# Current Status of Graphene Transistors

Max C. Lemme

Department of Physics, Harvard University, 17 Oxford Street, Cambridge, MA02138, USA,
lemme@fas.harvard.edu

**Keywords**: graphene, transistor, MOSFET, mobility, non-volatile memory

**Abstract**

This paper reviews the current status of graphene transistors as potential supplement to silicon CMOS technology. A short overview of graphene manufacturing and metrology methods is followed by an introduction of macroscopic graphene field effect transistors (FETs). The absence of an energy band gap is shown to result in severe shortcomings for logic applications. Possibilities to engineer a band gap in graphene FETs including quantum confinement in graphene Nanoribbons (GNRs) and electrically or substrate induced asymmetry in double and multi layer graphene are discussed. Graphene FETs are shown to be of interest for analog radio frequency applications. Finally, novel switching mechanisms in graphene transistors are briefly introduced that could lead to future memory devices.

**Introduction**

Graphene has attracted enormous research interest since its experimental discovery in 2004 [1, 2]. It consists of carbon atoms arranged in a 2-dimensional honeycomb crystal lattice with a bond length of 1.42 Å [3]. A schematic of a single graphene layer is shown in Fig. 1a, including "armchair" and "zig-zag" edges, named after their characteristic appearance on the atomic scale. The carbon atoms in the graphene lattice are $sp^2$ hybridized and three of the four valence electrons participate in the bonds to their next neighbors (σ–bonds). The schematic in Fig. 1b shows these in green (color online). The fourth π-electron orbital is oriented perpendicular to the sheet and delocalized (Fig. 1b, red).

The graphene lattice is made up of two equivalent carbon sublattices A and B, which lead to crystal symmetry. As a consequence, the charge carriers can be described by the Dirac equation [4], i.e. the band structure of graphene exhibits a linear dispersion relation for charge carriers, with momentum $k$ proportional to energy $E$. Finally, the energy bands associated with the sublattices intersect at zero energy $E$ resulting in a semi-metal with no band gap ($E_g = 0$ eV). A schematic of the band structure in the vicinity of k = 0 including the Fermi level $E_F$ is shown in Fig. 1c.

Charge carriers in graphene possess a very small effective mass [5], and hence graphene shows extremely attractive material properties relevant to electronic devices. These include carrier mobilities of up to 15000 $cm^2$/Vs for graphene on $SiO_2$ [5], 27000 $cm^2$/Vs for epitaxial graphene [6] and 200000 $cm^2$/Vs for suspended graphene [7-9]. In addition, high current carrying capability exceeding $1x10^8$ A/$cm^2$ [10], high thermal conductivity [11, 12], high transparency [13] and mechanical stability [14] have been reported. While similar promising properties have been reported for carbon nanotubes (CNTs), the fact that graphene sheets can be processed with conventional CMOS-technology is potentially a huge advantage over CNTs. Despite the enthusiasm over the discovery of graphene, however, research is still at an early stage. In this review we therefore discuss the potential of graphene for electronic applications based on experimental data available to date. A short introduction to standard graphene fabrication and detection methods is followed by an overview of the state-of-the-art in graphene metal oxide semiconductor field effect transistors (MOSFETs). We then briefly discuss graphene transistors for high frequency operation und present various non-classic graphene switches.

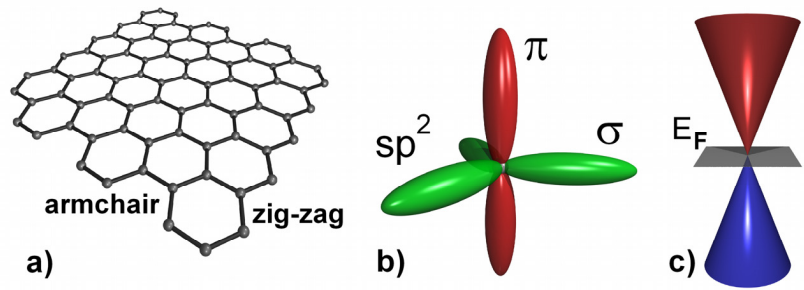

**Figure 1**: a) Schematic of a graphene crystallite with characteristic armchair and zig-zag edges. b) Schematic of electron σ– and π–orbitals of one carbon atom in graphene. c) Band diagram of graphene at k = 0.

**Graphene Fabrication**

There are currently three established fabrication methods for graphene: mechanical exfoliation, epitaxial growth from silicon carbide (SiC) substrates and chemical vapor deposition of hydrocarbons on reactive nickel or transition-metal-carbide surfaces.

    **Exfoliation**. Novoselov et al. have introduced a manual cleaving process of graphite, frequently called "mechanical exfoliation", to obtain single and few layer graphene [1, 4]. This process makes use of adhesive tape to pull graphene films off a graphite crystal. These are then thinned down by further strips of tape and finally rubbed against appropriate surfaces like silicon dioxide on silicon, which leaves randomly sized and distributed flakes on the surface. When observed through an optical microscope, single and few layer graphene flakes add to the optical path compared to the bare wafer. If a proper thickness of silicon dioxide is chosen, the resultant visible contrast is sufficient to identify even single graphene layers [15-18]. Fig. 2a shows the result of a contrast simulation of a single graphene layer on $SiO_2$. In this simulation, the contrast is plotted for a range of wavelengths and $SiO_2$ thicknesses. In the visible range, $SiO_2$ films of ~90 nm and ~300 nm result in high contrast and have hence been widely used as substrates. This pragmatic, low-cost method has enabled researchers to conduct a wide variety of fundamental physics and engineering experiments, even though it can not be considered a controlled process in terms of industrial exploitation. An example of typical graphene flakes on an oxidized silicon wafer is shown in Fig. 2b. While a trained person can distinguish single- from few layer graphene by "naked eye" with high fidelity, Raman spectroscopy has become the method of choice when it comes to scientific proof of single layers [19-22]. Both G and 2D Raman peaks at around 1580 $cm^{-1}$ and 2700 $cm^{-1}$ change in shape, position and intensity with the number of graphene layers. This is demonstrated in Fig. 1c, where Raman spectra for single- and few-layer graphene as well as graphite are plotted. Note that the spectra have been separated for ease of viewing by moving them along the y-axes. The baseline intensity is identical for each measurement.

    **Epitaxial Graphene form Silicon Carbide**. Berger and de Heer have pioneered an epitaxial approach to fabricate graphene from silicon carbide substrates [2, 6, 23]. During the process, silicon is thermally desorbed at temperatures between 1250°c and 1450°C. This process

can surely be classified more controllable and industrially relevant when compared to mechanical cleaving. In fact, it has been shown that graphene transistors can be manufactured from epitaxial graphene on a wafer scale [24]. Similar to exfoliated graphene, it has been demonstrated that single epitaxial graphene layers can be identified by Raman spectroscopy [25]. In addition Raman spectroscopy revealed that these layers are compressively strained [25, 26]. A major disadvantage of epitaxial graphene is the extremely high cost of production type SiC wafers, their limited size compared to silicon wafers, and the high processing temperatures well above current CMOS limits. With these pros and cons, it remains to be seen whether epitaxial graphene will find its way into future nanoelectronics applications.

**Chemical Vapor Deposition**. A promising large area deposition method is currently explored in the form of chemical vapor deposition on metallic surfaces like nickel [27-30], ruthenium [31] and others (see references in [32]). These CVD approaches rely on dissolving carbon into the metal substrates and then forcing it to precipitate out by cooling. Another approach is to grow graphene films directly on iridium [32, 33] or copper [34]. Several methods of transferring the CVD graphene films onto relevant substrates have been suggested, including the use of disposable PMMA or PDMS films. After transfer, Raman spectroscopy has been used to verify single layers, and even devices have been fabricated with typical graphene properties [29, 30, 34].

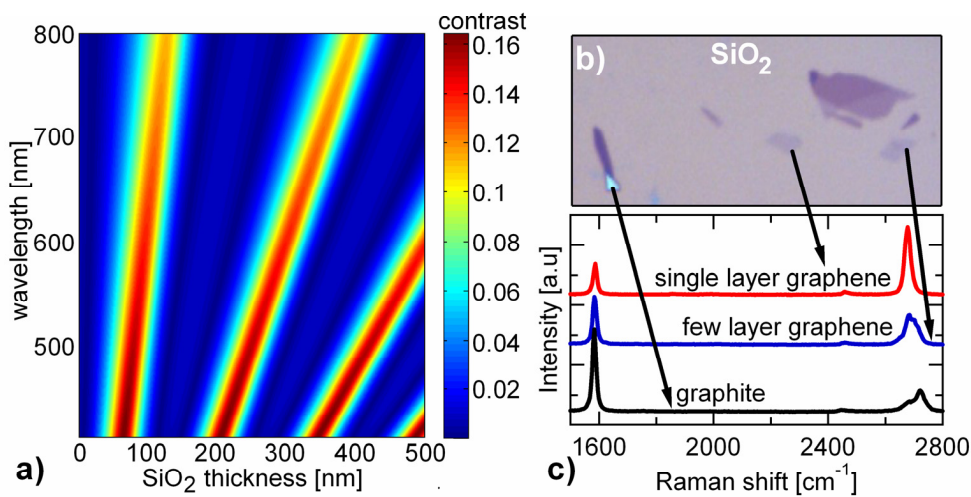

**Figure 2**: a) Contrast simulation of single layer graphene on a silicon dioxide film on silicon. Blue represents areas of no contrast, whereas red represents the maximum contrast. b) Optical micrograph of macroscopic graphene crystallites of various thicknesses on an $SiO_2$/ Si substrate. c) Raman spectra of three representative flakes of single- and few layer graphene and graphite.

## Macroscopic Graphene Field Effect Transistors

The most straightforward device application of graphene may seem to be as a replacement channel material for silicon MOSFETs. Fig. 3a shows a schematic of such a graphene field effect transistor (FET), including a top gate electrode, gate dielectric and source and drain metal contacts. Fig. 3b shows a top view optical micrograph of a macroscopic graphene field effect transistor on silicon dioxide (image modified and reproduced with permission from ECS Transactions, 11(6) (2007) [35]. Copyright 2007, The Electrochemical Society). The fabrication of graphene FETs follows standard silicon process technology once the graphene is deposited and identified. This includes the use of photo- or ebeam lithography, reactive ion etching and thin film deposition for the gate insulators and contacts. Details of typical fabrication processes are described in references [36-38].

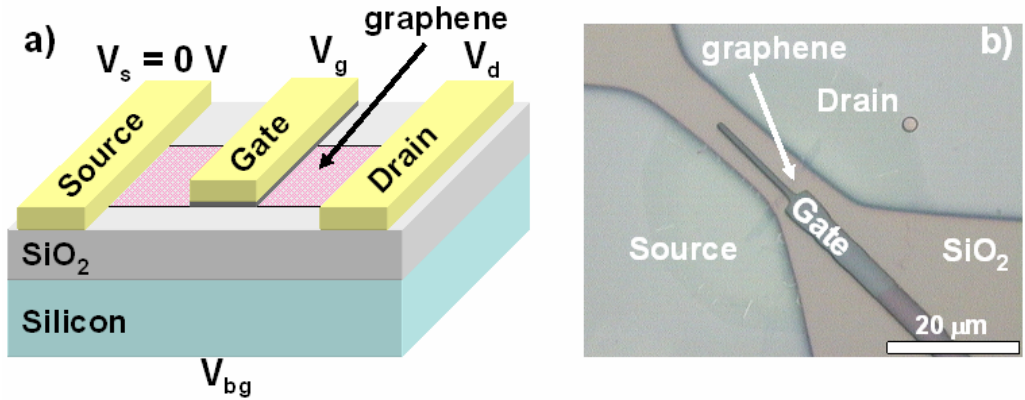

**Figure 3**: a) Schematic cross section and b) optical top-view micrograph of a graphene field effect transistor.

The transfer characteristics (here: drain current $I_d$ vs. back gate voltage $V_{bg}$) of a typical graphene transistor is shown in Fig. 4a. It reveals a major drawback of macroscopic graphene MOSFETs: the absence of an energy band gap ($E_g = 0$ eV) severely limits the current modulation in the graphene FET and, in addition, leads to ambipolar behavior. In fact, the best current modulation reported to date has been about 30, measured at cryogenic temperatures [1]. Furthermore, in conjunction with randomly distributed oxide charges the zero band gap leads to a finite minimum charge density even without any applied gate voltage [39]. Consequently, macroscopic graphene transistors conduct substantial current even at their point of minimum conductance (also referred to as Dirac point or charge neutrality point), preventing their application as a silicon MOSFET replacement in future CMOS-type logic circuits.

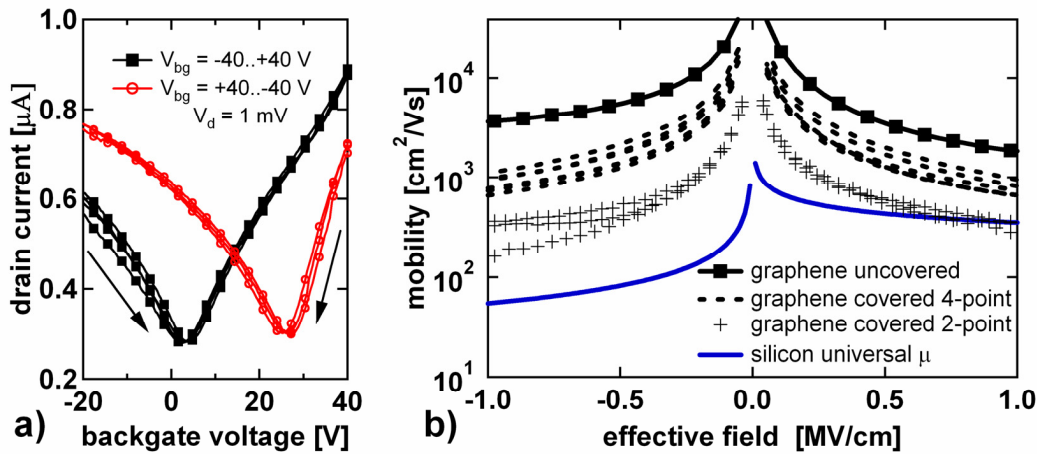

**Figure 4**: a) Drain current versus back gate voltage of a graphene FET. Changing the sweep direction results in considerable hysteresis of $\Delta V = 22$ V. b) Mobility versus electric field in graphene FETs. Covering graphene with a gate insulator leads to mobility reduction. Contacts have a considerable influence on graphene FETs. Universal mobility of silicon included as reference (after Takagi [40]).

Fig. 4a further shows hysteresis as the gate voltage is swept from negative to positive direction and vice versa. This typical behavior occurs despite measuring in vacuum conditions (P = 5x10$^{-3}$ mbar) and is a strong indicator of charge traps near the graphene / insulator interface. While suspended graphene measured in ultra high vacuum conditions has been shown to have mobilities exceeding 200000 cm$^2$/Vs, realistic graphene FETs are limited in performance by substrates and top gates. Nonetheless, the carrier mobilities in top gated devices exceed those of silicon and are typically on the order of several hundred to a thousand cm$^2$/Vs, even though graphene / insulator interfaces have not at all been optimized yet [35-38, 41, 42]. Fig. 4b shows electron and hole mobilities extracted from several top gated devices, both in 2-point and by 4-point probe configuration. (graph modified and reproduced with permission from ECS Transactions, 11(6) (2007) [35]. Copyright 2007, The Electrochemical Society).

**Graphene Nanoribbon Transistors**

A potential method to create a band gap in graphene is to cut it into narrow ribbons of less than a few tens of nanometers (graphene nanoribbons, GNRs). However, GNRs must be divided into two sub-types as indicated in Fig. 1: armchair and zig-zag edge terminated ribbons. Both types of GNR may be semiconducting or semimetallic. In armchair ribbons, the transition from 2D graphene to 1D GNRs leads to quantum confinement and a bandgap that is roughly inversely proportional to the nanoribbon width ($E_g \sim 1/W$) according to simulations [43, 44]. The precise value of the band gap is further predicted to depend on the number N of carbon atoms across the ribbon [43-46]. This is demonstrated in Fig. 5a, where the simulated density of states (DOS) versus energy for three different hydrogen-terminated armchair GNRs with N = 11, 12 and 13 atoms across the GNR width is shown [47, 48]. While the GNR with N = 11 is semimetallic, the ribbons with 12 and 13 atoms are semiconducting (generally armchair ribbons are semimetallic at N = 3m − 1, where m is an integer [49]). In hydrogen-terminated zig-zag GNRs, however, the situation is more complicated. It has been predicted by Nakada et al. that localized edge states near the Fermi level lead to semimetallic behavior, regardless of the number of carbon atoms [43]. On the other hand, Son et al. have calculated *ab initio* that edge magnetization causes a staggered sublattice potential on the graphene lattice that induces a band gap [45]. Finally, GNRs with other chiral orientation have been considered, including a mix of edges along a ribbon, adding to the complexity of this option [50-56]. In summary, the simulated results for any form of GNRs should be regarded with care, as they typically share an optimistic assumption of well controlled termination of dangling bonds. In reality, however, there is very likely a great variety of chemical groups terminating the edge atoms of a single graphene nanoribbon. A first detailed discussion has been recently published to address these issues [57], but it is probably reasonable to consider the nature of "real life" zig-zag GNRs an open question at this point in time.

The predicted presence of a band gap in specific GNRs has been experimentally confirmed. First evidence was reported by Han et al. [58] and Chen et al. [59], where GNRs were structured by e-beam lithography and etched in oxygen plasma with minimum widths of ~ 20 nm. The band gaps of these GNRs where in the range of ~30 meV and resulted in field effect transistors with $I_{on}/I_{off}$ ratios of about 3 orders of magnitude at low temperatures (1.7 – 4 K), reduced to a ratio of ~ 10 at room temperature. These investigations support theoretical predictions that sub-10 nm GNRs are required for true field effect transistor action at room temperature. More importantly, the experiments revealed a band gap regardless of the chiral orientation of the GNRs [58]. This latter result was attributed to a strong influence of edge states, which dominates over the chirality dependency of the band structure. To date, two examples of sub-10 nm GNRs have been shown experimentally. Ponomarenko et al. have fabricated GNRs with a minimum width of about 1 nm and a band gap of about 500 meV using e-beam lithography and repeated, careful overetching [60]. The resulting transistors consequently switched off at room temperature to "no measurable conductance". An alternative fabrication process for GNRs has been presented by Li et al. [61]. Here, graphene ribbons were solution derived from graphite by thermal exfoliation, sonification and centrifugation. The resulting solution was dispersed onto substrates and GNRs were identified with an atomic force microscope (AFM). The resulting devices exhibited well behaved transistor action at room temperature with $I_{on}/I_{off}$ ratios of more than $10^6$ [61, 62]. Finally, unzipping of carbon nanotubes by etching or sputtering has been experimentally shown to result in GNRs [63, 64]. Interestingly, all GNR transistors in these studies were semiconducting, even though both armchair and zig-zag orientations were believed to be present. The experimental $I_{on}/I_{off}$ ratios reported to date are summarized in Fig. 5b. While they clearly support theoretical predictions and show promise for graphene nanoribbon electronics, they also show an urgent need for further research in this field: Statistical data is obviously scarce and the discrepancies between theory and experiment have to be addressed. The necessity of controllable sub-10nm feature sizes and great uncertainties in chirality control as well as edge state definition remain tremendous challenges towards future industrial applicability. To this end, a recently developed technique, helium ion beam microscopy, has been shown to have potential for precise nanopatterning of graphene [65, 66].

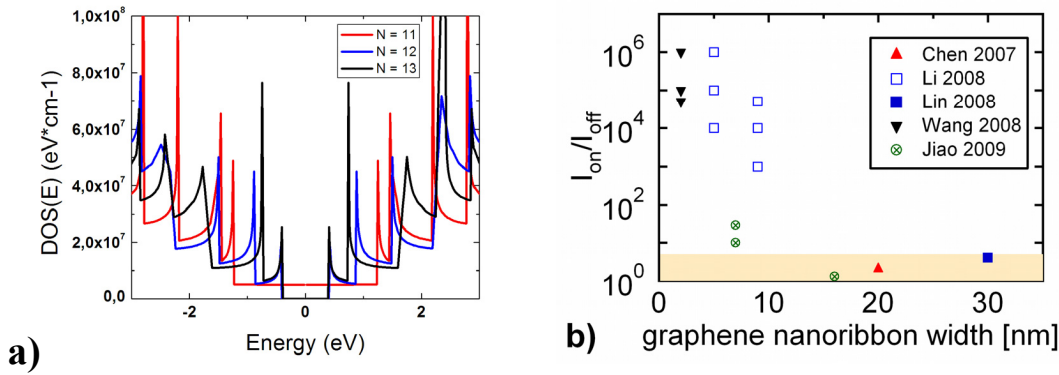

**Figure 5:** a) Simulated density of states (DOS) versus energy of hydrogen terminated armchair graphene nanoribbons (GNRs) for various numbers of electrons N across the ribbon [47]. b) Experimental $I_{on}/I_{off}$ ratios versus GNR width taken from literature. None of the GNRs reported thus far has shown metallic behavior. The lower colored part of the graph indicates $I_{on}/I_{off}$ values of typical macroscopic graphene FETs.

**Bilayer Graphene and Substrate Effects**

A viable approach to obtain a band gap in graphene is to break its symmetry. McCann proposed that macroscopic double- or bilayer graphene films would display a band gap if a transverse electric field was applied to them to cause layer asymmetry [67]. His calculations predict a roughly linear dependence of the band gap on the carrier density n, with each $10^{12}$ cm$^{-2}$ adding about 10 meV to the gap. This prediction was experimentally confirmed by Ohta et al. on bilayer graphene films on silicon carbide (SiC) through angle-resolved photoemission spectroscopy (ARPES) [68]. In their work, they used potassium doping to modify the carrier density in the graphene, which lead to changes in the electronic band gap. Oostinga et al. took this approach a step further by applying an electrostatic field through a top gate electrode. Their bilayer graphene device showed a band gap in the range of a few meV, demonstrated with low temperature measurements at 4K [69].

A related concept has been demonstrated by Zhou et al., who report a much more significant bandgap in single layer graphene on SiC of 260 meV, again obtained by ARPES measurements [70]. As the number of graphene layers increases, the band gap was found to decrease. Here, the gap is attributed to a broken A,B sublattice symmetry in the graphene, caused by a buffer layer between the SiC crystal and the graphene [71]. While this approach seems to indicate the feasibility of macroscopic graphene transistors with high $I_{on}/I_{off}$ ratios at room temperature, there are no experimental reports of such devices on single layer graphene on SiC to date. Even an extensive study by Kedzierski et al. [24], who investigated a statistically relevant number of several hundreds of graphene transistors on SiC, did not reveal a considerable band gap in any of the devices. While mobility values exceeding those of silicon transistors were achieved, all devices showed typical "zero-gap" graphene characteristics similar to Fig. 3c. In summary, electrically induced band gaps in bilayer graphene have been observed, but their small absolute value prevents an application in room temperature field effect transistors. While a larger substrate induced bandgap has been observed by ARPES for single layer graphene on SiC, this result remains to be confirmed in an actual electronic device.

**RF Transistors**

The discussed lack of a band gap at room temperature in macroscopic graphene FETs makes them unsuitable for logic applications. For radio-frequency (RF) analog applications, on the other hand, a high on-off ratio is desirable but not mandatory. Instead, most important for good RF performance is a FET channel with excellent carrier transport properties (high mobility and maximum velocity) [72], combined with a small scale length, which improves strongly as the channel material thickness is reduced [73]. As graphene fulfils these requirements, graphene RF FETs have recently been investigated, both made from exfoliated [42, 74] and from epitaxial graphene [75]. The reported cut off frequencies are plotted as a function of gate lengths in Fig. 6. The maximum cut off frequency reported was $f_T = 26$ GHz, but the mobility values reported were far from ideal, and higher $f_T$ can be expected for optimized devices with shorter gate lengths. This once more emphasizes the need for graphene / insulator interface engineering.

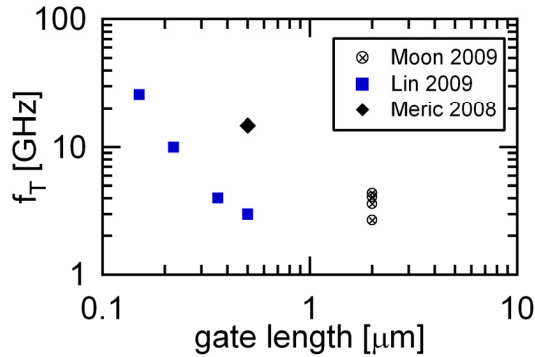

**Figure 6**: Summary of published graphene RF transistor cut off frequencies.

**Non-Conventional Graphene Switches**

A number of concepts for (non-volatile) graphene switches have emerged that operate on mechanisms other than the classic semiconductor field effect. Even though a thorough review is beyond the scope of this paper, they are briefly introduced in this section.

    A first concept are graphene/graphene-oxide (GO) Schottky barrier MOSFETs [76], where semiconducting GO acts as the transistor channel.

    Another approach suggests atomic scale graphene switches that are based on creating nanoscale gaps by electric fields in graphene films [77]. These physical gaps are reversibly opened and closed by breaking and re-forming the carbon atomic chains in the graphene.

Chemical surface modification affects strongly the electronic band structure of graphene[78]. We reversibly modified the drain current electrostatically in a graphene FET similar to Fig. 3a by controlled chemisorption [79].

Ferroelectric gating has been shown to electrostatically dope graphene and change the drain currents in a non-volatile way [80].

While these early concepts are far from mature, they nevertheless demonstrate the potential of graphene for nanoelectronics applications that might not be anticipated today.

**Conclusions**

We have reviewed the potential of graphene-based field effect transistors to supplement or substitute existing silicon CMOS technology. While macroscopic graphene transistors are not suitable for logic application due to the lack of an energy band gap, graphene RF transistors seem promising and feasible. Graphene nanoribbons, on the other hand, show extreme promise as a straightforward CMOS compatible approach, but their extreme sensitivity on an atomic level to both geometric and edge termination variations may well render their application impossible. Finally, recent discoveries of non-classic switching mechanisms may eventually lead to a co-integration of graphene into silicon technology, even though details are obviously not yet clear today. In addition to these device related issues, a major roadblock for the application of graphene is the unavailability of a large area, CMOS compatible deposition technique. Fortunately, as CVD and related methods are being explored, many device related questions are being addressed today using the existing manufacturing methods of exfoliation and graphene epitaxy. These insights will be transferable as soon as industrially more relevant technologies become available.

**Acknowledgment**
M.C. Lemme thanks T.J. Echtermeyer, B. Szafranek, J.R. Williams and S. Nakaharai for fruitful discussions and gratefully acknowledges the support of the Alexander von Humboldt foundation through a Feodor Lynen Research Fellowship. Simulation services for results presented here were provided by the Network for Computational Nanotechnology (NCN) at nanoHUB.org.